\documentclass[12pt]{article}
\usepackage{microtype} \DisableLigatures{encoding = *, family = * }
\usepackage{graphicx,multirow}
\usepackage{amsmath}
\usepackage{amsfonts,amssymb}
\usepackage{natbib}
\bibpunct{(}{)}{;}{a}{,}{,}
\usepackage[bookmarksnumbered=true, pdfauthor={Wen-Long Jin}, breaklinks=true]{hyperref}

\newcommand{\commentout}[1]{}

\newcommand{\ba}{\begin{array}}
	\newcommand{\ea}{\end{array}}
\newcommand{\bc}{\begin{center}}
	\newcommand{\ec}{\end{center}}
\newcommand{\bdm}{\begin{displaymath}}
\newcommand{\edm}{\end{displaymath}}
\newcommand{\bds} {\begin{description}}
	\newcommand{\eds} {\end{description}}%17Apr01
\newcommand{\ben}{\begin{enumerate}}
	\newcommand{\een}{\end{enumerate}}
\newcommand{\beq}{\begin{equation}}
\newcommand{\eeq}{\end{equation}}
\newcommand{\bfg} {\begin{figure}[htbp]}
	\newcommand{\efg} {\end{figure}}%Nov 5,99
\newcommand{\bi} {\begin {itemize}}
\newcommand{\ei} {\end {itemize}}
\newcommand{\bsl} {\begin{slide}[8.8in,6.7in]}
	\newcommand{\esl} {\end{slide}}
\newcommand{\bsq}{\begin{subequations}}
	\newcommand{\esq}{\end{subequations}}
\newcommand{\bss} {\begin{slide*}[9.3in,6.7in]}
	\newcommand{\ess} {\end{slide*}}
\newcommand{\btb} {\begin {table}}
\newcommand{\etb} {\end {table}}%Nov 10,99
\newcommand{\bvb}{\begin{verbatim}}
	\newcommand{\evb}{\end{verbatim}}

\newcommand{\reff}[1] {{{Figure \ref {#1}}}}
\newcommand{\refe}[1] {{(\ref {#1})}}%Nov 5
\newcommand{\reft}[1] {{{\textbf{Table} \ref {#1}}}}
\def\pmb#1{\setbox0=\hbox{$#1$}%
	\kern-.025em\copy0\kern-\wd0
	\kern.05em\copy0\kern-\wd0
	\kern-.025em\raise.0433em\box0 }

\def\eop{{\hfill $\blacksquare$}}%17Apr01
\newtheorem{theorem}{Theorem}[section]%17Apr01
\newtheorem{lemma}[theorem]{Lemma}%17Apr01
\begin {document}
\author{Wen-Long Jin\footnote{Department of Civil and Environmental Engineering, California Institute for Telecommunications and Information Technology, Institute of Transportation Studies, 4000 Anteater Instruction and Research Bldg, University of California, Irvine, CA 92697-3600. Tel: 949-824-1672. Fax: 949-824-8385. Email: wjin@uci.edu. Corresponding author}}

\title{Provably safe and human-like car-following behaviors: Part 2. A parsimonious multi-phase model with projected braking} %20250428Mon11:09PDT@MBP2316

\maketitle
\begin{abstract}
Ensuring safe and human-like trajectory planning for automated vehicles amidst real-world uncertainties remains a critical challenge. While existing car-following models often struggle to consistently provide rigorous safety proofs alongside human-like acceleration and deceleration patterns, we introduce a novel multi-phase projection-based car-following model. This model is designed to balance safety and performance by incorporating bounded acceleration and deceleration rates while emulating key human driving principles.
Building upon a foundation of fundamental driving principles and a multi-phase dynamical systems analysis (detailed in Part 1 of this study \citep{jin2025WA20-02_Part1}), we first highlight the limitations of extending standard models like Newell's with simple bounded deceleration. Inspired by human drivers' anticipatory behavior, we mathematically define and analyze projected braking profiles for both leader and follower vehicles, establishing safety criteria and new phase definitions based on the projected braking lead-vehicle problem.
The proposed parsimonious model combines an extended Newell's model for nominal driving with a new control law for scenarios requiring projected braking. Using speed-spacing phase plane analysis, we provide rigorous mathematical proofs of the model's adherence to defined safe and human-like driving principles, including collision-free operation, bounded deceleration, and acceptable safe stopping distance, under reasonable initial conditions. Numerical simulations validate the model's superior performance in achieving both safety and human-like braking profiles for the stationary lead-vehicle problem. Finally, we discuss the model's implications and future research directions.
\end{abstract}


\section{Introduction}
The planning stage in driving, for both human-driven and automated vehicles, is paramount for ensuring safety amidst a complex and dynamic environment \citep[][Chapter 6]{arkin1998behavior}. It bridges perception with action, strategizing trajectories based on myriad factors including safety, comfort, vehicle limits, and traffic laws. While advancements in sensing and AI are significant \citep{shalev2017formal}, the core challenge in planning lies in developing driving models that are not only computationally efficient but also provably safe and exhibit human-like behaviors, attributes often found to be conflicting in existing car-following literature.

Part 1 of this study \citep{jin2025WA20-02_Part1} introduced a multi-phase dynamical systems analysis framework and applied it to standard car-following models (Newell's, Intelligent Driver Model, Gipps), revealing their limitations in consistently satisfying a comprehensive set of safe and human-like driving principles, particularly concerning braking dynamics and absolute safety guarantees under bounded deceleration. For instance, naive extensions of simple models like Newell's to include bounded deceleration can lead to safety violations, the Gipps model is ill-defined when the follower is within the safe jam spacing and inconsistent with the observed fundamental diagram, and the Intelligent Driver model exhibits backward traveling and excessive stopping distance. This motivates the need for novel car-following models that are both parsimonious and can verifiably ensure safety while replicating nuanced human driving patterns.

Human drivers inherently employ projection, anticipating potential future actions of lead vehicles, especially during braking, and planning their own maneuvers accordingly. This proactive behavior is crucial for maintaining safety margins. While the concept of safe stopping distance has been explored \citep{gipps1981bcf, shalev2017formal}, a rigorous, parsimonious car-following model that explicitly incorporates projected braking dynamics into a multi-phase control structure and other human-like behaviors with provable safety guarantees has been lacking.

This article (Part 2) proposes such a solution: a parsimonious multi-phase projection-based car-following model, building upon the foundational principles and analytical framework established in Part 1. Our contributions are threefold:
\begin{enumerate}
    \item We mathematically formalize the concept of projected braking for both leader and follower, defining safety criteria and new driving phases (nominal driving and comfort braking) based on these projections and their relationship to minimum and comfort jam spacings.
    \item We develop a novel multi-phase projection-based car-following model that integrates an extended Newell's model for nominal driving conditions with a new control law specifically designed for scenarios demanding projected braking, ensuring adherence to bounded acceleration and deceleration.
    \item We provide rigorous mathematical proofs demonstrating that this model is collision-free and adheres to other critical safe and human-like driving principles under realistic initial conditions, offering a significant improvement over existing models analyzed in Part 1.
\end{enumerate}
The proposed model's parsimony aims for easier calibration and potential for real-world learning, while its multi-phase projection-based nature allows for both human-like nominal driving and robustly safe emergency responses. Numerical simulations validate the theoretical analysis and compare the model's braking profiles and safety performance against alternatives. We conclude by discussing implications and future extensions for this approach in the pursuit of advanced vehicular automation.

The remainder of this article is organized as follows. Section 2 briefly revisits key definitions and the principles of safe and human-like driving behaviors pertinent to this work, as well as the extended Newell's car-following model, drawing from Part 1. Section 3 establishes the foundation for our projection-based car-following model, analyzing the projected braking lead-vehicle problem, formulating projection-based safety principles, and defining new phases. Section 4 introduces our multi-phase projection-based car-following model, presenting mathematical proofs of its adherence to safe and human-like driving principles. Section 5 presents numerical simulations and analysis, including the model's performance in the stationary lead-vehicle problem and its braking profile characteristics. Section 6 concludes the study, discussing potential future research directions and the implications of our findings.

\section{Brief review of variables, behavioral principles, and extended Newell's car-following model}

This section provides a concise overview of essential car-following variables, key behavioral principles for safe and human-like driving, and the extended Newell's car-following model that will serve as a component of our proposed multi-phase model. For a comprehensive discussion, derivation, and detailed analysis of these concepts and standard car-following models, the reader is referred to Part 1 of this study \citep{jin2025WA20-02_Part1}. A list of essential notations used in this article is given in \reft{t:notations_part2}.

\btb
\centering
\caption{Essential Notation and Description for Part 2}
\begin{tabular}{|c|p{10cm}|}
\hline
Notation        & Description                                       \\
\hline
$B(t)$ & Safe stopping distance\\
$V(\cdot)$ & Speed-density relation $v=V(k)$\\
$X(t)$, $X_L(t)$ & Location of follower, leader at time $t$              \\
$a(t)$        & Acceleration rate of the follower at time $t$     \\
$k(t)$ & Density ($k(t)=\frac 1{z(t)}$)\\
$q(t)$ & Flow-rate\\
$t$& Time\\
$t'$& Projected time\\
$v(t)$, $v_L(t)$ & Speed of follower, leader at time $t$                \\
$v_*(t)$        & Equilibrium speed of the follower at $t$  \\
$z(t)$        & Spacing of the follower ($X_L(t)-X(t)$)               \\
\hline
$\Delta t$ & Time-step size\\
$\alpha$        & Comfort acceleration bound of the follower            \\
$\beta$         & Comfort deceleration bound of the follower            \\
$\beta_L$       & Projected deceleration rate of the leader \\ % Added
$\epsilon$      & Infinitesimal time step ($=\Delta t$ in discrete version)\\
$\kappa$ & Jam density ($=\frac 1{\zeta}$)\\
$\mu$           & Speed limit                                       \\
$\tau$          & Minimum time gap (for nominal driving, e.g., Newell's)            \\
$\tau'$       & Reaction time of follower during braking \\ % Clarified
$\zeta$         & Comfort jam spacing                \\
$\zeta'$        & Minimum jam spacing                        \\
$\zeta_c$ & Critical spacing ($=\tau \mu+\zeta$)\\
\hline
$\tilde B(t)$ & Projected braking distance of follower\\
$\tilde X(t',t)$, $\tilde X_L(t',t)$ & Projected location of follower, leader at $t'$ during braking (from current time $t$)  \\
$\tilde X(t)$, $\tilde X_L(t)$ & Projected stopping location of follower, leader (from current time $t$)  \\
$\tilde z(t)$ & Projected final spacing when vehicles stop ($\tilde X_L(t)-\tilde X(t)$)     \\
$\tilde z(t',t)$ & Projected spacing at future projected time $t'$ (based on state at $t$)  \\
\hline
$\Phi(v,v_L)$      & Projected stopping comfort jam spacing function    \\ % Simplified notation
$\Phi'(v,v_L)$      & Projected stopping minimum jam spacing function    \\ % Simplified notation
$\tilde \zeta(t)$& Projected comfort jam spacing (depending on $v_L(t)$)\\
$\tilde \zeta'(t)$&Projected minimum jam spacing (depending on $v_L(t)$) \\

\hline
\end{tabular}\label{t:notations_part2}
\etb

\subsection{Recap of core variables and behavioral principles}

We consider a follower vehicle at location $X(t)$ with speed $v(t)$ and acceleration $a(t)$, trailing a leader vehicle at $X_L(t)$ with speed $v_L(t)$. The spacing is $z(t) = X_L(t) - X(t)$. The dynamics are typically described by:
\bsq
\begin{align}
X(t+\epsilon)&=X(t)+\epsilon v(t+\epsilon), \label{eq:kin_X_p2}\\
v(t+\epsilon)&=v(t)+\epsilon a(t),\label{eq:kin_v_p2}
\end{align}
\esq
where $\epsilon$ is an infinitesimal time step.

The design and evaluation of car-following models are guided by a set of behavioral principles detailed in \citep{jin2025WA20-02_Part1}. These include:
\begin{itemize}
    \item \textbf{Zeroth-Order Collision-free Principles:} Maintaining a minimum jam spacing ($z(t) \ge \zeta'$) and, for comfort, a larger comfort jam spacing ($z(t) \ge \zeta$).
    \item \textbf{First-Order Operational Constraints:} Ensuring forward travel ($v(t) \ge 0$), adherence to speed limits ($v(t+\epsilon) \le \mu$), and maintaining a minimum time gap ($\tau(t) \ge \tau$, implying $v(t+\epsilon) \le (z(t)-\zeta)/\tau$).
    \item \textbf{Second-Order Acceleration Constraints:} Operating within comfort acceleration bounds ($a(t) \le \alpha(1-v(t)/\mu)$) and comfort deceleration bounds ($a(t) \ge -\beta$). Additionally, the concept of a safe stopping distance ($B(t) = v(t)\tau' + v^2(t)/(2\beta)$), considering reaction time $\tau'$, is crucial for safety.
    \item \textbf{Driver Objectives and Macroscopic Properties:} Aiming to maximize speed where safe and ensuring consistency with observed fundamental diagrams in steady states.
\end{itemize}
This study prioritizes adherence to these principles, particularly ensuring collision-free trajectories and bounded, human-like acceleration/deceleration.

\subsection{The extended Newell's car-following model as a nominal driving component}

Part 1 of this study \citep{jin2025WA20-02_Part1} analyzed Newell's simplified car-following model \citep{newell2002carfollowing} and its extensions. Newell's model, in its basic form, determines the planned speed $v(t+\epsilon)$ as the minimum of the speed limit $\mu$ and a speed dictated by the comfort spacing $\zeta$ and minimum time gap $\tau$:
\begin{align}
v_*(t) = \min\left\{\mu, \frac{z(t)-\zeta}{\tau}\right\}, \quad v(t+\epsilon) = v_*(t). \label{eq:Newell_basic_p2}
\end{align}
This model ensures comfort jam spacing and produces a triangular fundamental diagram but allows unbounded acceleration/deceleration.

To address these limitations, extensions incorporating bounded acceleration ($\alpha$) and bounded deceleration ($\beta$) were examined. The Bounded-Acceleration Newell (BA-Newell) model successfully limits acceleration but not deceleration. The Bounded-Deceleration-and-Acceleration Newell (BDA-Newell) model, which applies both bounds:
\begin{align}
a(t) = \max\left\{-\beta, \min\left\{\alpha\left(1-\frac{v(t)}{\mu}\right), \frac{v_*(t)-v(t)}{\epsilon}\right\}\right\}. \label{eq:BDA_Newell_a_p2}
\end{align}
In Part 1, five phases in the speed-spacing phase plane were analyzed, and it was shown that this model potentially violates the minimum jam spacing and forward travel principles when braking from high speeds towards a stationary leader. It was also shown that the initial spacing cannot be smaller than the comfort jam spacing, which could be the case when vehicles stop at signalized intersections or change lanes.

For the multi-phase projection-based model proposed in this paper, we utilize the BDA-Newell model, as defined by \refe{eq:BDA_Newell_a_p2} and the subsequent speed update via \refe{eq:kin_v_p2}, specifically for the nominal driving phase where conditions are deemed sufficiently and proactive projected braking is not immediately required. This provides a parsimonious basis for comfortable acceleration and cruising behavior. The limitations of this BDA-Newell component in ensuring safety under all braking scenarios motivate the introduction of the projection-based braking phase developed in the subsequent sections of this paper.

\section{Foundation of projection in car-following: problems, concepts, and definitions}
To balance the behavioral principles in the preceding section that ensure safety and human-like driving behaviors, a following vehicle must proactively prepare for the possibility of the lead vehicle braking and plan its trajectory accordingly. This requires the follower to peer into the future, both in terms of time and space, extending beyond the leader's present location. We call this behavior projection. While this shares conceptual similarities with anticipatory car-following models in the literature \citep{treiber2006delays}, our approach differs in several important aspects. Existing anticipatory models typically incorporate leader velocity differences or simple extrapolations, whereas our projection-based framework explicitly calculates complete braking trajectories under comfort braking scenarios to ensure provable safety guarantees while maintaining human-like behavior. To mathematically formalize this concept, we introduce the projected braking lead-vehicle problem. Subsequently, we establish projected collision-free principles and delineate phases grounded in this projection-based framework.

It is important to distinguish this projection-based approach from traditional safety distance models. While both aim to ensure collision-free operation, traditional safety distance approaches like those used in the Gipps model focus primarily on maintaining a buffer based on the final stopping position. In contrast, our projection framework explicitly models the complete temporal and spatial evolution of both vehicles' trajectories during braking. This comprehensive approach ensures safety not just at the final stopping positions but throughout the entire braking process. Furthermore, it allows us to overcome the limitations of traditional safety distance models, such as ill-definedness for certain spacing ranges and inconsistencies between fundamental diagram parameters and safe stopping requirements (see Section 5 of Part 1 \citep{jin2025WA20-02_Part1}).

\subsection{Projected braking lead-vehicle problem}

\bfg\bc
\includegraphics[width=5.5in]{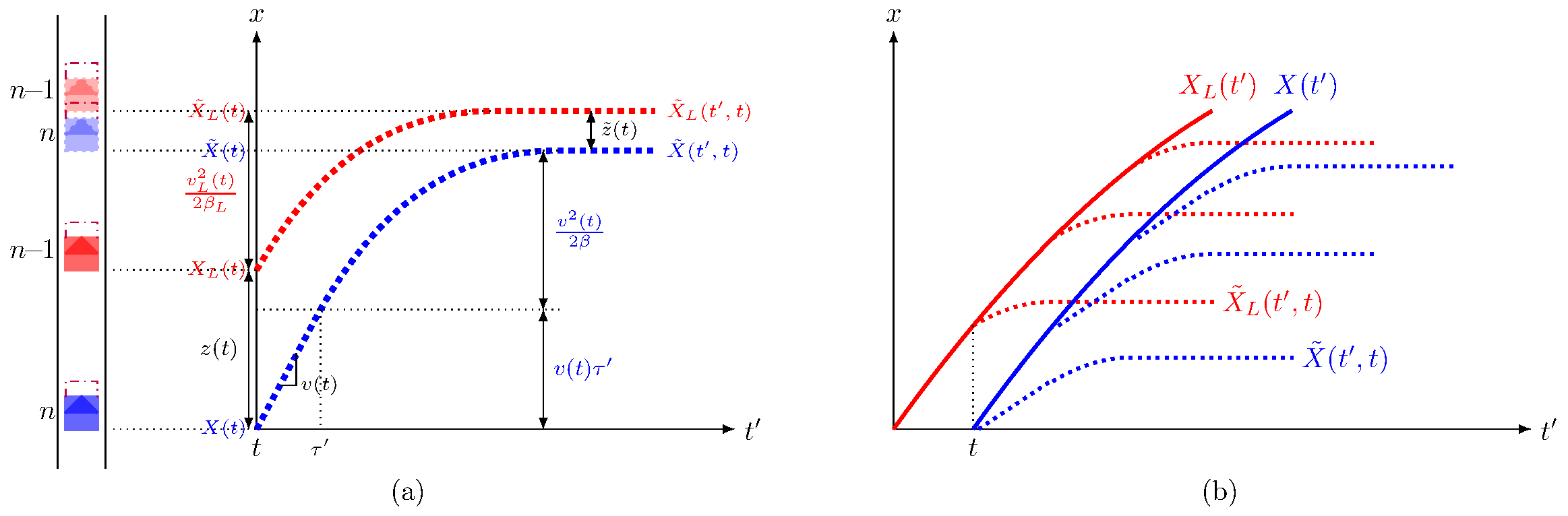}\caption{Projected braking lead-vehicle problem}\label{fig:BLVP-illustration}
\ec\efg

At each time instant, we assume that the follower (the blue cars in \reff{fig:BLVP-illustration}(a)) plans its trajectory by anticipating the leader (the red cars in \reff{fig:BLVP-illustration}(a)) to brake and come to a complete stop. In other words, the follower continuously solves what we refer to as the projected braking lead-vehicle problem at every time step. Up to time $t$, the follower observes the leader's trajectory as $X_L(t)$; subsequently, the follower projects a braking procedure, and the projected braking trajectory for $t' \geq t$ is represented as $\tilde X_L(t',t)$, where $t'$ is the projected time. For simplicity, the leader is projected to maintain a constant deceleration rate $\beta_L$, and the projected braking trajectory $\tilde X_L(t',t)$ is defined as follows:
\begin{align}\label{eqn:projected-leader-trajectory}
\tilde  X_L(t',t) &=
\begin{cases}
  X_L(t'),  & \text{if } t'\leq t; \\
  X_L(t)+v_L(t) (t'-t)-\frac 12 \beta_L (t'-t)^2, & \text{if }  t<t'\leq t+\frac{v_L(t)}{\beta_L};\\
  \tilde X_L(t)\equiv  X_L(t)+\frac{v^2_L(t)}{2\beta_L},&\text{if }  t'> t+\frac{v_L(t)}{\beta_L}.
\end{cases}
\end{align}
Here $\tilde X_L(t,t) = X_L(t)$, and the leader stops at $\tilde X_L(t)$.
The projected braking trajectory is parabolic before stopping for future times. This is illustrated by the red, thick, dotted curve in \reff{fig:BLVP-illustration}(a).

Immediately at $t$, when the leader is projected to brake, the follower also needs to plan for stopping. Here we assume that the follower has a reaction time of $\tau'$, during which it travels at a constant speed of $v(t)$. After that, it brakes at a constant deceleration rate $\beta$. Thus, the projected braking trajectory of the following the follower, $\tilde X(t',t)$, is defined as follows:
\begin{align}\label{eqn:projected-follower-trajectory}
\tilde  X(t',t) &=
\begin{cases}
  X(t'),  & \text{if } t'\leq t; \\
  X(t)+v(t) (t'-t), & \text{if } t<t'\leq t+\tau';\\ 
  X(t)+v(t) (t'-t)-\frac 12 \beta (t'-t-\tau')^2, & \text{if }  t+\tau'<t'\leq t+\tau'+\frac{v(t)}{\beta};\\
  \tilde X(t)\equiv  X(t)+v(t) \tau'+\frac{v^2(t)}{2\beta},&\text{if }  t'> t+\tau'+\frac{v(t)}{\beta}.
\end{cases}
\end{align}
Here $\tilde X(t,t) = X(t)$,   and the follower stops at $\tilde X(t)$.
The follower's projected braking trajectory is illustrated by the blue, thick, dotted curve in \reff{fig:BLVP-illustration}(a).

In \reff{fig:BLVP-illustration}(b), we illustrate the actual trajectories, denoted as $X_L(t')$ and $X(t')$, alongside several projected trajectories, $\tilde X_L(t',t)$ and $\tilde X(t',t)$, all presented in the $(t',x)$ plane. These projected trajectories form surfaces in the $(t',t,x)$ space, derived from the given actual trajectories. The figure shows that the projected trajectories can deviate significantly from the actual ones. Since future leader behavior is unpredictable, it's a safer approach for the follower to rely on projected braking scenarios. However, when the projected deceleration rate, denoted as $\beta_L$, becomes infinite, the leader is projected to stop instantaneously, simulating an abrupt accident scenario. Conversely, when the projected deceleration rate is zero, it implies that the leader is anticipated to maintain a constant cruising speed.

We represent the projected spacing between the two vehicles as $\tilde z(t',t)$:
\begin{align}
\tilde z(t',t)&=\tilde X_L(t',t)-\tilde X(t',t),
\end{align}
and the projected stopping spacing between both stopping vehicles as $\tilde z(t)$: 
\begin{align}
\tilde z(t)&=\tilde X_L(t)-\tilde X(t)=z(t)-v(t) \tau'+\frac{v^2_L(t)}{2\beta_L}-\frac{v^2(t)}{2\beta},
\end{align}
which is also illustrated in \reff{fig:BLVP-illustration}(a).
Regarding the relationship between these spacings, the following lemma is presented, with its detailed proof provided in the Appendix.

\begin{lemma} \label{lemma:projected-spacing} The projected spacing at time $t\geq t'$ attains its minimum either at time $t$ or when both vehicles come to a complete stop under the following situations.
\bsq\label{projected-spacing-lemma-conditions}
\begin{enumerate}
\item The follower's deceleration rate, $\beta$, is not greater than the leader's, $\beta_L$; i.e.,
\begin{align}
\beta&\leq \beta_L. \label{projected-spacing-lemma-condition-1}
\end{align}
\item Or, the follower decelerates more rapidly and stops later than the leader; i.e.,
\begin{align}
\beta&>\beta_L, \text{and } \tau'+\frac{v(t)}{\beta} \geq \frac{v_L(t)}{\beta_L}. \label{projected-spacing-lemma-condition-2}
\end{align}
\item Or, the follower decelerates more rapidly, is slower at $t$ and not faster at $t+\tau'$; i.e.,
\begin{align}
\beta&>\beta_L, v(t)< v_L(t), \text{and } \tau'+\frac{v(t)}{\beta_L} \geq \frac{v_L(t)}{\beta_L}. \label{projected-spacing-lemma-condition-3}
\end{align}
\een
\esq
In other words, if the conditions defined in \refe{projected-spacing-lemma-conditions} are satisfied, we have (for $t'\geq t$)
\begin{align}
 \tilde z(t',t) &\geq \min\{z(t), \tilde z(t)\}. \label{relation:projected-spacing}
\end{align}
In contrast, if the follower decelerates more rapidly, is not slower at $t$, and stops earlier than the leader; i.e.,
\begin{align}
\beta&>\beta_L, v(t)\geq v_L(t), \text{and } \tau'+\frac{v(t)}{\beta} < \frac{v_L(t)}{\beta_L}. \label{projected-spacing-lemma-conditions-opposite-1}
\end{align}
then \refe{relation:projected-spacing} is violated and 
\begin{align}
\min_{t'\geq t} \tilde z(t',t) &< \min\{z(t), \tilde z(t)\}.\label{opposite-relation:projected-spacing}
\end{align}
If the follower decelerates more rapidly, is slower at $t$, stops earlier, and is faster at $t+\tau'$ than the leader; i.e.,
\begin{align}
\beta&>\beta_L, v(t)< v_L(t),  \tau'+\frac{v(t)}{\beta} < \frac{v_L(t)}{\beta_L}, \nonumber\\
&\text{and } v(t)> v_L(t)- \beta_L\tau', \label{projected-spacing-lemma-conditions-opposite-2}
\end{align}
then both \refe{relation:projected-spacing} and \refe{opposite-relation:projected-spacing} could be valid.

\end{lemma}

\subsection{Projected safety and projection-based phases}

With the projected braking trajectories of both the leader and follower, we can extend the collision-free principles in Section 2 to a projected time $t'$. For the projected braking comfort jam spacing principle, the projected spacing should not be smaller than the comfort jam spacing:
\begin{align}
& \tilde z(t',t) \geq \zeta. \label{PB-CJS-principle}
\end{align}
For the projected braking  minimum jam spacing principle, the projected spacing should not be smaller than the vehicle length plus the minimum safety cushion:
\begin{align}
& \tilde z(t',t) \geq \zeta'. \label{PB-MJS-principle}
\end{align}

Both projected braking comfort and minimum jam spacing principles imply the following safety principles when both vehicles are projected to stop:
\begin{enumerate}
\item Projected stopping comfort jam spacing principle:
\begin{align}
& \tilde z(t) \geq \zeta. \label{PS-CJS-principle}
\end{align}
\item Projected stopping minimum jam spacing principle:
\begin{align}
& \tilde z(t) \geq \zeta'. \label{PS-MJS-principle}
\end{align}
\end{enumerate}

From Lemma \ref{lemma:projected-spacing}, we have the following theorem, whose proof is straightforward and therefore omitted.
\begin{theorem}
If and only if \refe{relation:projected-spacing} is satisfied, the projected braking comfort jam spacing principle is equivalent to
comfort jam spacing plus the projected stopping comfort jam spacing principle, 
and the projected braking minimum jam spacing principle is equivalent to the minimum jam spacing principle 
plus the projected stopping minimum jam spacing principle. In these cases, projected braking safety at any $t'>t$ is equivalent to actual safety at $t$ plus projected stopping safety at $t'=\infty$. A sufficient condition for the equivalence is \refe{projected-spacing-lemma-conditions}.
\end{theorem}

When \refe{projected-spacing-lemma-conditions} is satisfied, the speed-spacing plane can be subdivided into the following phases:
\ben
\item In the nominal driving phase, the state satisfies both comfort jam spacing and projected stopping comfort jam spacing principles:
\bsq \label{PB-nominal-phase}
\begin{align}
z(t)&\geq \zeta,\\
z(t)&\geq \Phi(v(t);v_L(t)),
\end{align}
where the projected stopping comfort jam spacing condition \refe{PS-CJS-principle} is defined by a quadratic function, referred to as the
projected stopping comfort jam spacing function:
\begin{align}
\Phi(v(t);v_L(t))&=\tilde \zeta(t)+v(t)\tau'+\frac{v^2(t)}{2\beta}, \label{PS-CSS-curve}
\end{align}
and the projected comfort jam spacing is
\begin{align}
\tilde \zeta(t)&=\zeta- \frac{v^2_L(t)}{2\beta_L}. 
\end{align}
\esq
For the stationary lead-vehicle problem, $\tilde \zeta(t)=\zeta$ at any time $t$.

\item In the comfort braking phase, the state satisfies both the minimum jam spacing and projected stopping minimum jam spacing principles but not projected braking comfort jam spacing principle:
\bsq \label{PB-CB-phase}
\begin{align}
z(t)&\geq \zeta',\\
z(t)&\geq \Phi'(v(t);v_L(t)),
\end{align}
where the projected stopping minimum jam spacing condition \refe{PS-MJS-principle} is defined by another quadratic function, 
referred to as the projected stopping minimum jam spacing function:
\begin{align}
\Phi'(v(t);v_L(t))&=\tilde \zeta'(t)+v(t)\tau''+\frac{v^2(t)}{2\beta},  \label{PS-MSS-curve}
\end{align}
where we simply set the reaction time $\tau''=\frac 12 \tau',$\footnote{The following discussions are valid for any $\tau''\in[0,\frac 12 \tau']$.} 
and the projected minimum jam spacing is
\begin{align}
\tilde \zeta'(t)&=\zeta'- \frac{v^2_L(t)}{2\beta_L}. 
\end{align}
\esq
For the stationary lead-vehicle problem, $\tilde \zeta'(t)=\zeta'$ at any time $t$.

\item In the emergency braking phase, when the state satisfies the minimum jam spacing principle but not the projected stopping minimum jam spacing principle:
\bsq \label{emergency-phase}
\begin{align}
z(t)&\geq \zeta',\\
z(t)&< \Phi'(v(t);v_L(t)).
\end{align}
\esq

\item In the collision phase, when the state does not satify the minimum jam spacing principle:
\begin{align}
z(t)&< \zeta'. \label{collision-phase}
\end{align}
\een

\bfg\bc \includegraphics[width=4in]{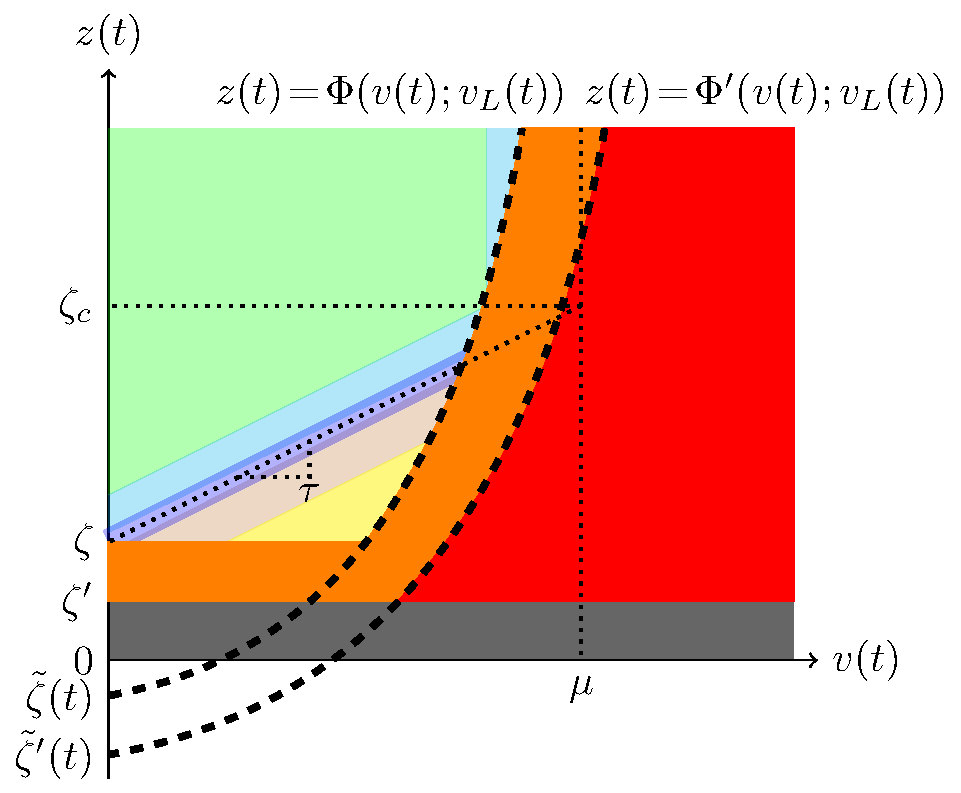} \caption{Projection-based phases in the $(v(t),z(t))$ plane}\label{fig:new-BD-phase} \ec\efg

\reff{fig:new-BD-phase} illustrates phases based on projected braking. In this figure, the two quadratic functions defined in equations \refe{PS-CSS-curve} and \refe{PS-MSS-curve} are depicted as thick dashed curves. The top-left region corresponds to the nominal driving phase, the middle orange region to the comfort braking phase, the right red region to the emergency braking phase, and the bottom black region to the collision phase. The two quadratic curves remain separate, since $\Phi(v(t);v_L(t))-\Phi'(v(t);v_L(t))=\zeta-\zeta'>0$. These phases change with the leader's speed $v_L(t)$.

\section{A multi-phase projection-based car-following model with assured safety and bounded deceleration}

In this section, we present a multi-phase car-following model based on the projected braking concepts in the preceding section. We will demonstrate that the model has assured safety and  bounded deceleration for initial states in both the 
projected braking comfort jam spacing and projected braking minimum jam spacing phases.

\subsection{A multi-phase projection-based car-following model}
Building upon the projection-based framework introduced in the preceding section, our aim is to develop a model that prevents traffic states from transitioning from 
projected braking comfort jam spacing and projected braking minimum jam spacing phases into the emergency phase. The emergency phase is characterized by a 
lack of assured safety according to the projected stopping minimum jam spacing principle. Additionally, our model should ensure that vehicles do not travel in reverse.

Intuitively, if the initial state falls within the projected braking comfort jam spacing phase, where 
\begin{align}
z(t)\geq \max\{\zeta,\Phi(v(t);v_L(t))\}, \label{nominal-CB-boundary}
\end{align}
it is positioned at a considerable distance from the emergency braking phase, given that 
\begin{align*}
\max\{\zeta,\Phi(v(t);v_L(t))\}>\max\{\zeta',\Phi'(v(t);v_L(t))\}.
\end{align*}
In such cases, we can employ the BDA-Newell model as described in \refe{eq:BDA_Newell_a_p2}. Since both $z(t)$ and $\Phi'(v(t);v_L(t))$ change at limited rates, the subsequent state remains outside the emergency braking phase. Furthermore, when the initial speed of the follower is zero, it's important to note that the BDA-Newell model prescribes a non-negative acceleration rate due to the condition $z(t) \geq \zeta$. Consequently, the subsequent speed remains non-negative, ensuring that vehicles consistently move forward.

Nevertheless, when an initial state resides within the comfort braking phase, the application of the BDA-Newell model does not ensure that the subsequent state will avoid entry into the emergency phase or prevent vehicles from moving backwards. This issue is exemplified by the stationary lead-vehicle problem as follows. 
\bi
\item First, if the initial speed is $v(t)=0$, and the initial spacing is $z(t)=\zeta'$, the BDA-Newell model, as previously discussed in Section 2, would prescribe a negative speed for the follower. 
\item Second, in a scenario where the initial speed is $v(t)=\mu$, and the initial spacing is $z(t)=\Phi'(\mu;0)$, employing the BDA-Newell model results in a subsequent state where the speed remains $\mu$ but the spacing diminishes: $z(t+\epsilon)<z(t)=\Phi'(\mu;0)$. Consequently, this state transitions into the emergency braking phase.
\ei
Thus, the necessity for a novel driving model emerges, and our approach is to assume that the follower enters the projected braking process and derive the acceleration rate accordingly. In this scenario, the projected braking distance for the follower is defined as:
\begin{align}
 \tilde B(t)&=z(t)-\Phi'(v(t);v_L(t))+\frac{v^2(t)}{2\beta}.\label{def:B}
\end{align}
Clearly,
\begin{align}
 \tilde B(t)&=z(t)- \frac 12 v(t)\tau'- \zeta'+\frac{v^2_L(t)}{2\beta_L},\label{def:B2}
\end{align}
which is the available distance for the follower to come to a complete stop with a minimum jam spacing in the event that the leader stops at a constant deceleration rate. Then the corresponding acceleration rate is given by
\begin{align}
a(t)&=-\frac{v^2(t)}{2 \tilde B(t)}. \label{def:PB-MSS-a}
\end{align}

Combining the previously presented models for both phases, we arrive at the comprehensive multi-phase projection-based car-following model:
\bsq\label{MPP-CFM}
\begin{align}
&a(t)= \begin{cases}
\max\{-\beta, \min\{\alpha(1-\frac{v(t)}{\mu}), \frac{v_*(t)-v(t)}{\epsilon}\}\}, \text{if }z(t)\geq \max\{\zeta,\Phi(v(t);v_L(t))\};&\\
-\frac{v^2(t)}{2  \tilde B(t)}, \text{ if }\max\{\zeta',\Phi'(v(t);v_L(t))\} \leq z(t)< \max\{\zeta,\Phi(v(t);v_L(t))\};&
\end{cases}\label{MPP-a}\\
&v(t+\epsilon)=v(t)+\epsilon a(t),\label{MPP-v}\\
&X(t+\epsilon)=X(t)+\epsilon v(t+\epsilon).\label{MPP-x}
\end{align}
\esq

Referring to the phase diagram in  \reff{fig:new-BD-phase}, the nominal driving phase can be subdivided into five distinct sub-phases following the BDA-Newell model's classifications: bounded acceleration (green), equilibrium acceleration (cyan), equilibrium cruising (blue), equilibrium deceleration (brown), and bounded deceleration (yellow), consistent with those illustrated in Figure 4 of \citep{jin2025WA20-02_Part1}. 

We can observe that the new car-following model has distinct spatial and temporal properties. It is localized with respect to vehicles, as it only requires information about the immediate leader-follower pair. However, it is non-local with respect to time and space, as it projects trajectories far into the future to determine current actions. The time horizon for these projected braking scenarios is relatively long, typically on the order of 20 seconds. Despite this extended planning horizon, the execution stage remains local, with the follower computing and applying a specific acceleration rate at each discrete time step $t$. This approach exemplifies the principle of global planning with local execution, where safety is ensured through comprehensive future projections while maintaining computational efficiency through localized implementation.

\subsection{Behavioral analysis}

In this subsection,  we rigorously establish the analytical properties of the new car-following model in accordance with the behavioral principles outlined in Section 2. Here, we make the assumption that $\beta\geq \beta_L$. This means that under normal, safe driving conditions, we should plan our driving trajectory based on the assumption that the follower can brake more aggressively than the leader.

First, we establish that the model adheres to the principles of bounded acceleration and bounded deceleration in the following theorem, whose proof is in the Appendix.
\begin{theorem}\label{thm:BD}
In both the nominal driving and comfort braking phases, the proposed car-following model maintains bounded acceleration and deceleration rates, ensuring: $-\beta\leq a(t) \leq \alpha$. 
\end{theorem}

We now proceed to establish the adherence of the model to the forward traveling principle in the following theorem, whose proof is in the Appendix.
\begin{theorem}\label{thm:FT}
In both the nominal driving and comfort braking phases, the follower's speed in the proposed car-following model remains non-negative: $v(t)\geq 0$. Thus, the new model adheres to the forward traveling principle. 
\end{theorem}

Furthermore, we establish the model's consistent adherence to the minimum jam spacing principle. First, all states within both the nominal driving and comfort braking phases conform to the following minimum jam spacing inequality:
\begin{align*}
z(t)\geq \max\{\zeta',\Phi'(v(t);v_L(t))\},
\end{align*}
which can be equivalently expressed as:
\begin{align}
\min\{z(t)-\zeta',z(t)-\Phi'(v(t);v_L(t))\}\geq 0. \label{inequality:CSS}
\end{align}
This inequality ensures the minimum jam spacing principle within the model's behavior. Then we have the following theorem, whose proof is in the Appendix.

\begin{theorem}\label{thm:CSS}
In both the nominal driving and comfort braking phases, the proposed car-following model adheres to the minimum jam spacing principle. This means that if the minimum jam spacing inequality in \refe{inequality:CSS} is satisfied at a given time step, it remains valid in subsequent steps. It's important to note that states may transition between the nominal driving and comfort braking phases.
\end{theorem}

For the new car-following model within a homogeneous traffic stream, where all vehicles share identical characteristics, the steady state within the nominal driving phase yields the same speed-density relationship that for Newell's simplified car-following model. In the comfort braking phase, the steady state occurs when $v_L(t)=v(t)=0$, and $\kappa<k\leq \frac{1}{\zeta'}$. Consequently, we arrive at the following theorem, the proof of which is straightforward and omitted.

\begin{theorem}\label{thm:FD}  The proposed car-following model adheres to the fundamental diagram principle, and the fundamental diagram is an extended version of the triangular one in \citep{jin2025WA20-02_Part1} ($k\in [0,\frac 1{\zeta'}]$):
\bsq \label{extended-triangualr-FD}
\begin{align}
v&=\max\{0,\min\{\mu, \frac 1\tau (\frac 1k-\frac 1\kappa)\}\},\\
q&=\max\{0,\min\{\mu k, \frac 1\tau (1-\frac k\kappa)\}\}.
\end{align}
\esq

\end{theorem}

The extended triangular fundamental diagram is depicted in \reff{fig:extended-triangularFD}. The maximum density is $\frac{1}{\zeta'}$. For instance, with values $\zeta = 7$ m and $\zeta' = 5$ m, the jam density $\kappa=\frac 1 {\zeta}$ is approximately 70\% of the maximum density. This observation aligns with real-world data, where the jam occupancy is typically around 0.7, rather than reaching a full 1. Readers can refer to Figure 13 in the study by \citep{yan2018automatic} and its associated references for more details.

\bfg\bc \includegraphics[width=4in]{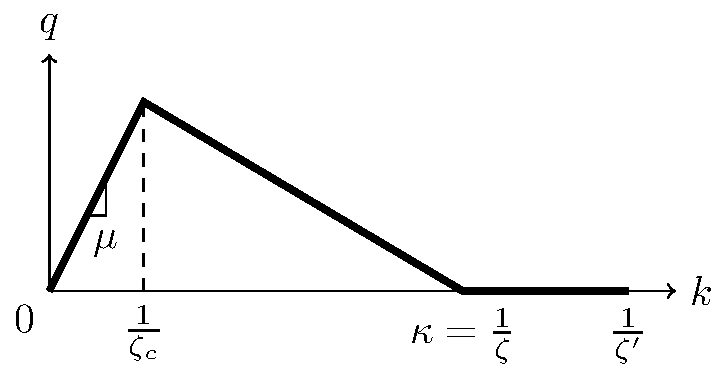} \caption{Extended triangular fundamental diagram for the multi-phase projection-based car-following model \refe{MPP-CFM}}\label{fig:extended-triangularFD} \ec\efg

\section{Application to stationary lead-vehicle problem}
In this section, we solve the stationary lead-vehicle problem to obtain the braking profile, first analytically and then numerically.

\subsection{Analytical solutions of the stationary lead-vehicle problem}

In this subsection, we solve the stationary lead-vehicle problem with  $v_L(t)=0$ for all $t$. Initially, the follower is far away and traveling at a speed $v(0)>0$. We assume that the vehicle enters the comfort braking phase from the nominal driving phase at $t=0$. From \refe{nominal-CB-boundary}, we have that $\Phi(v(0);0)>\zeta$ and 
\begin{align}
z(0)&=\Phi(v(0);0)=\zeta+v(0) \tau' +\frac{v^2(0)}{2\beta}. \label{SLVP-initial}
\end{align}
For $t>0$, the vehicle's dynamics are governed by \refe{def:PB-MSS-a}, yielding:
\bsq
\begin{align}
\frac{d}{dt} z(t)&=- v(t),\label{dX-equation}\\
\frac{d}{dt} v(t)&=-\frac{v^2(t)}{2\tilde B(t)}, \label{dv-equation}
\end{align}
\esq
where the projected braking distance $\tilde B(t)$ at $t\geq 0$ is
\begin{align}
\tilde B(t)&=z(t)- \frac 12 v(t) \tau' -\zeta', \label{def-Bt}
\end{align}
with $\tilde B(0)= \zeta-\zeta' +\frac{\tau'}2 v(0)+\frac{v^2(0)}{2\beta} $. 

From the above equations, we obtain the jerk from the acceleration rate $a(t) = \frac{d}{dt}v(t)$ as 
\begin{align*}
\frac{d}{dt}a(t) &= \frac{\tau'v^4(t)}{8\tilde{B}^3(t)}=-\tau'\frac{a^3(t)}{v^2(t)}.
\end{align*}
To find $a$ as a function of $v$, we use $\frac{d}{dt}a(t)= \frac{da}{dv} \cdot \frac{dv}{dt} = \frac{da}{dv} \cdot a(t)$, giving 
\begin{align*}
\frac{da}{dv} &= -\tau' \frac{a^2(t)}{v^2(t)}.
\end{align*}
This separable equation becomes $\frac{da}{a^2} = -\frac{\tau'}{v^2}dv$, with solution $a(v) = -\frac{v}{\tau' + Cv}$. With initial conditions $v(0) > 0$, $z(0) = \zeta + v(0)\tau' + \frac{v^2(0)}{2\beta}$, and $\zeta > \zeta'$, $\beta > 0$, we find $\tilde{B}(0) = \zeta - \zeta' +\frac{\tau'}2 v(0)+ \frac{v^2(0)}{2\beta}$, and $a(0)=-\frac{v^2(0)}{ 2(\zeta-\zeta')+\tau' v(0)+\frac{v^2(0)}{\beta}}$. The constant $C = \frac{2(\zeta - \zeta')}{v^2(0)}  + \frac{1}{\beta}$, giving the final expression 
\bsq\label{SLVP-solutions}
\begin{align}
a(v) = -\frac{v}{\tau' + (\frac{2(\zeta - \zeta')}{v^2(0)}  + \frac{1}{\beta})v}. \label{SLVP-a_v}
\end{align}
This expression for $a(v)$ allows us to analyze the vehicle's deceleration profile directly in the speed domain. Particularly, we have
\begin{align}
\frac{d}{dt}a(t) &=\frac{\tau'v}{\left(\tau' + \left(\frac{2(\zeta - \zeta')}{v^2(0)}  + \frac{1}{\beta}\right)v\right)^3},\\
\tilde B(v)&= \frac{\tau'}2 v + (\frac{\zeta - \zeta'}{v^2(0)}  + \frac{1}{2\beta}) v^2,\\
z(v)&= \zeta' + \tau'v + (\frac{(\zeta - \zeta')}{v^2(0)}  + \frac{1}{2\beta}) v^2.
\end{align}
\esq
The properties of the solutions are presented in the following theorem, whose proof is in the Appendix.

\begin{theorem}\label{theorem:SLVP-properties}
For the stationary lead-vehicle problem with $v_L(t)=0$ for all $t$, \refe{SLVP-solutions}, which is the solution of the multi-phase projection-based car-following model \refe{MPP-CFM} during the comfort braking phase, has the following properties:
\begin{enumerate}
\item[(i)] The vehicle always decelerates with $a(t)<0$ before stopping, and the deceleration is bounded by the comfort deceleration bound: $-a(t) \leq \beta $ for all $t \geq 0$.
\item[(ii)] The vehicle decelerates without traveling backward: $v(t) \geq 0$ for all $t \geq 0$.
\item[(iii)] No collision occurs: $z(t) \geq \zeta'$ for all $t \geq 0$.
\item[(iv)] The vehicle converges to a complete stop at the minimum jam spacing: $\lim_{t \to \infty} v(t) = 0$ and $\lim_{t \to \infty} z(t) = \zeta'$.
\item[(v)] From \refe{SLVP-initial}, the stopping distance equals $z(0)-\zeta=v(0)\tau'+\frac{v^2(0)}{2\beta}$, which satisfies the safe stopping distance principle in Section 2.
\end{enumerate}
\end{theorem}

\subsection{A numerical example}

For the numerical simulation based on \refe{MPP-CFM}, we initialized the values as follows: $v(0)=0$ m/s, $z(0)=2500$ m, and $v_L(t)=0$ m/s for all $t$. We use a time-step size of $\Delta t=0.001$ s, and introduce the following parameters: comfort jam spacing $\zeta=7$ m, minimum jam spacing $\zeta'=5$ m, minimum time gap $\tau=1.6$ s,  reaction time $\tau'=1$ s, desired speed $\mu=120$ km/h, comfort acceleration bound $\alpha=0.73$ m/s$^2$, and comfort deceleration bound $\beta=1.67$ m/s$^2$. We use the same initial conditions and parameters as those in the example in Figure 2 of \citep{treiber2000congested}, with an additional parameter $\tau'=1$.

\bfg
\centering
  \includegraphics[width=6in]{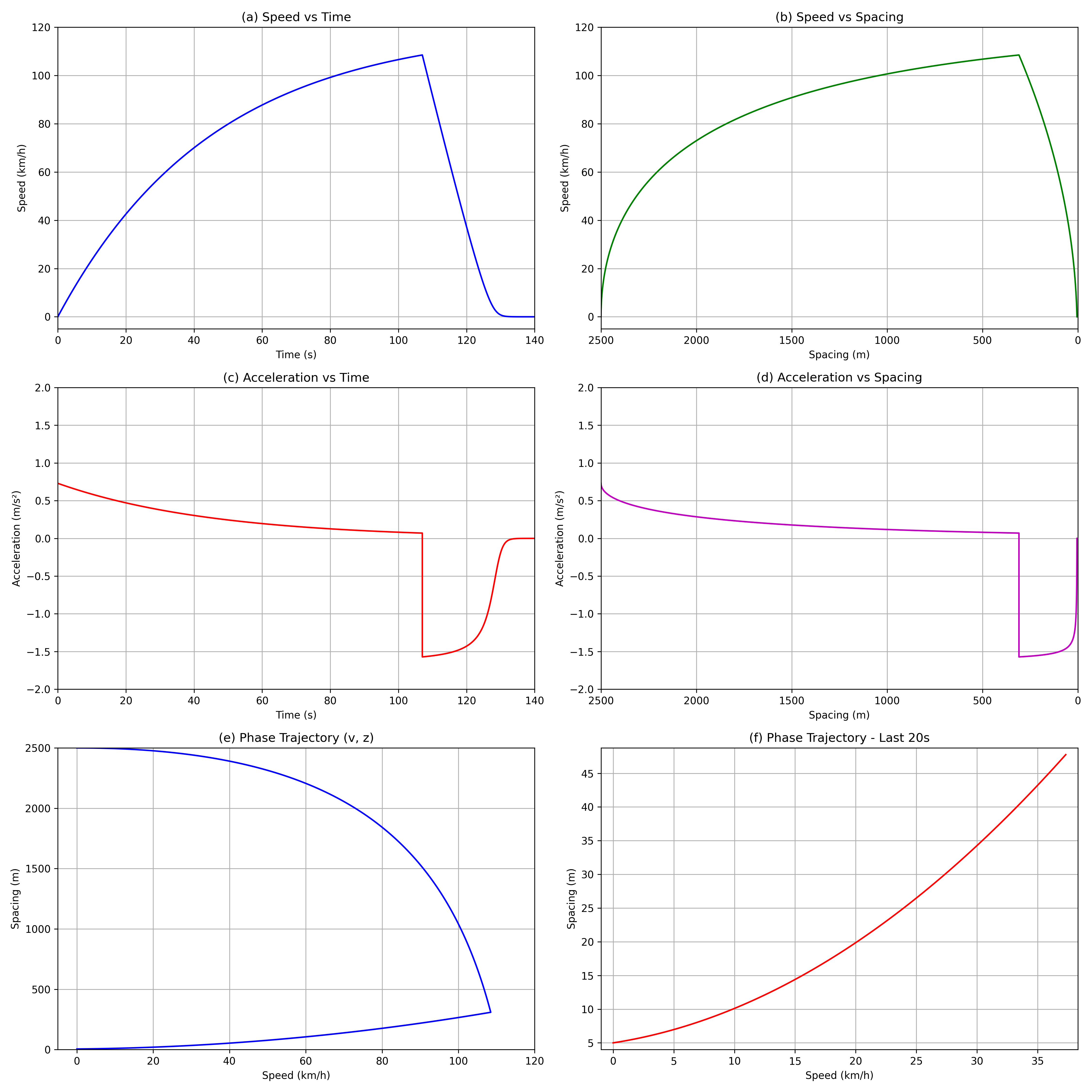}
  \caption{Replication of Figure 2 in \citep{treiber2000congested} with the multi-phase projection-based car-following model}
  \label{fig:MP_CF_140s}
\efg

In \reff{fig:MP_CF_140s}, the six panels show: (a) speed versus time, (b) speed versus spacing, (c) acceleration versus time, (d) acceleration versus spacing, (e) phase trajectory in the speed-spacing plane for the entire simulation, and (f) phase trajectory for the last 20 seconds.
From the figures we can see that: 
\bi
\item First, panel (f) confirms that the trajectory in the speed-spacing plane converges to the minimum jam spacing, and the speed is always non-negative.
\item Second, panels (a) and (b) show that the vehicle accelerates to the maximum speed of about 108 km/h with the nominal driving phase of the model, and then enters the comfort braking phase to brake to stop. The stopping distance is about 302 m, consistent with that given by the safe stopping distance principle.
\item Third, a closer look at panels (c) and (d) reveals that, when the vehicle switches from the first phase to the second, there is a sharp change in the acceleration rate, from a positive value to a minimum value of about -1.6 m/s$^2$. This suggests that the multi-phase projection-based car-following model has no bound on the jerk, which is a limitation.
\ei

These results verify the analytical results in the preceding subsection, and further reveal a limitation related to infinite jerks. These simulation results are close to those for the simplified Gipps model in Figure 7 in \citep{jin2025WA20-02_Part1}, except that the follower stops at the minimum jam spacing (5 m), not the comfort jam spacing (7 m).

\section{Conclusion}
This article introduced a novel multi-phase projection-based car-following model designed to ensure provable safety while exhibiting human-like driving behaviors and maintaining parsimony. Building upon the fundamental driving principles and the multi-phase dynamical systems analysis framework detailed in Part 1 of this study \citep{jin2025WA20-02_Part1}, which highlighted limitations in standard car-following models, we developed a model that addresses these shortcomings. Our approach incorporates a human-inspired projected braking mechanism, allowing the follower vehicle to proactively plan its trajectory based on anticipated leader braking.

The core contributions of this work (Part 2) include: first, the mathematical formalization of projected braking dynamics and the definition of new safety-critical phases based on these projections. Second, the development of the multi-phase projection-based model itself, which combines an extended Newell's model for nominal driving with a dedicated control law for scenarios requiring projected braking, ensuring adherence to bounded acceleration and deceleration. Third, and critically, we provided rigorous mathematical proofs demonstrating that this model is collision-free (maintaining minimum jam spacing) and adheres to the forward traveling principle and bounded control under reasonable initial conditions within its defined operational phases. Numerical simulations of the stationary lead-vehicle problem validated these theoretical guarantees and showcased the model's ability to produce human-like braking profiles while avoiding the pitfalls (e.g., excessive deceleration, backward travel, or collisions) identified in simpler extensions of Newell's model in Part 1.

Beyond its technical merits, the multi-phase projection-based car-following model achieves a remarkable balance between theoretical comprehensiveness and parametric parsimony. It introduces only two additional parameters beyond the BDA-Newell model ($\tau'$ and $\zeta'$), one more than the Intelligent Driver Model ($\tau'$), and two more than the Gipps model ($\tau$ and $\zeta'$). Despite this minimal parameter expansion, as demonstrated throughout Sections 4 and 5, the model successfully addresses all the limitations identified in these standard approaches while maintaining clear physical interpretability of each parameter. Based on our analyses and the extensive literature review, we conjecture that any car-following model capable of satisfying all the behavioral principles outlined in Section 2, up to the second-order derivatives (acceleration and deceleration), would require at least the seven parameters used in our formulation ($\zeta$, $\zeta'$, $\tau$, $\tau'$, $\mu$, $\alpha$, $\beta$), though the model structure could differ. The only remaining limitation is the lack of jerk constraints, which represents a potential avenue for future research while likely requiring additional parameters.

Our study makes several simplifying assumptions that warrant further discussion. For instance, we assume that the follower employs a single comfort deceleration bound ($\beta$) across all scenarios. In reality, human drivers may employ significantly different maximum deceleration rates in emergency situations versus normal driving \citep{durrani2021predicting,wood2021evaluating}. Such emergency scenarios, while important, fall beyond the scope of the current work, which focuses on establishing fundamental safety guarantees under normal operating conditions.

The proposed multi-phase projection-based car-following model represents a significant advancement toward driving systems that simultaneously satisfy three critical requirements: theoretical soundness with provable safety guarantees, human-like behavior patterns, and parametric parsimony enabling practical calibration. Building on this foundation, several promising research directions emerge:

\begin{itemize}
    \item \textbf{Expanded scenario analysis:} Extending the analytical solutions beyond the stationary lead-vehicle problem to scenarios including moving leaders, shockwave propagation through traffic streams, dilemma zone navigation at signalized intersections, and phantom jam formation and dissipation. The multi-phase framework established here provides a robust foundation for such analyses.
    
    \item \textbf{Empirical validation:} Calibrating and validating the model's seven core parameters ($\zeta$, $\zeta'$, $\tau$, $\tau'$, $\mu$, $\alpha$, $\beta$) against large-scale vehicle trajectory datasets to verify its real-world fidelity and potentially identify regional or demographic variations in parameter values.
    
    \item \textbf{Boundary condition exploration:} Investigating the model's behavior under conditions approaching or briefly exceeding the proven safe initial states, such as mild emergency situations. This includes developing extensions to handle vehicle heterogeneity, stochastic variations in driving behavior, and jerk limitations to enhance both comfort and realism.
    
    \item \textbf{Integration with broader driving systems:} Expanding the model to encompass complex maneuvers such as lane-changing, merging, and diverging, and incorporating it within comprehensive automated vehicle control architectures to evaluate its impact on safety, efficiency, and traffic flow stability.
\end{itemize}

While alternative approaches exist, such as control barrier functions \citep{ames2019control,alan2023control} that prioritize safety guarantees, our work reinforces the fundamental insight presented in Part 1 \citep{jin2025WA20-02_Part1}: safety and human-likeness are not competing objectives but rather complementary requirements for effective automated driving systems. As noted in our introduction, the gradual introduction of automated vehicles necessitates their coexistence with human-driven vehicles for the foreseeable future \citep{sperling2018three}. Human-like behavior is not merely an aesthetic concern but a critical safety feature, as it ensures predictability and compatibility with existing traffic flow patterns. Our model's balanced stopping distances align with human expectations and existing infrastructure design parameters, addressing the specific concern raised in Part 1 regarding traffic signals and dilemma zones \citep{gazis1960problem}. By maintaining stopping distances comparable to those of human drivers, our model ensures that automated vehicles can safely navigate signalized intersections designed with human reaction times and deceleration rates in mind. Furthermore, the model could inform driver education programs by establishing scientifically-grounded safe following practices, potentially improving safety even for non-automated vehicles. The explicit consideration of both comfort parameters ($\zeta$, $\alpha$) and safety parameters ($\zeta'$, $\beta$) reflects our nuanced approach to balancing performance and safety in mixed-autonomy environments.

Ultimately, the development of parsimonious, provably safe, and human-like car-following models represents an essential step not only for advancing vehicular automation technology but also for fostering public trust in autonomous driving systems through demonstrable safety guarantees and familiar behavior patterns.

\section*{Acknowledgments}
The author expresses sincere gratitude to the UC ITS Statewide Transportation Research Program (STRP) for their generous financial support. Special thanks are extended to Dr. Ximeng Fan for valuable discussions on the model, data, and related literature. The author gratefully acknowledges Prof. Magnus Egerstedt and Dr. Brooks Butler from UC Irvine for enlightening discussions on control barrier functions and their applications to vehicle safety. The author also thanks Dr. Zuduo Zheng of the University of Queensland and the anonymous reviewers of Transportation Research Part B for their constructive feedback on an earlier version of this manuscript. The author appreciates the assistance of AI language models during the iterative process of revising text and exploring concepts, which contributed to the efficiency of this work. Any presented results, perspectives, or errors are the sole responsibility of the author.

\section*{Declaration of generative AI and AI-assisted technologies in the writing process}
During the preparation of this work the author used Claude, ChatGPT, and Google Gemini in order to improve language clarity, enhance readability, and refine phrasing in specific sections. After using these tools, the author reviewed and edited all content thoroughly and takes full responsibility for the content of the publication.

\section*{Appendix: Proofs of lemmas and theorems}

\subsection*{Proof of Lemma \ref{lemma:projected-spacing}}
{\em Proof}. From the definition of $\tilde X_L(t',t)$ in \refe{eqn:projected-leader-trajectory} and $\tilde X(t',t)$ in \refe{eqn:projected-follower-trajectory}, we have the following projected speeds, which are both non-increasing piecewise linear functions of $t'$:
\begin{align*}
\frac{d}{dt'}\tilde X_L(t',t)  &=
\begin{cases}
 v_L(t)  - \beta_L (t'-t), & \text{if }  t<t'\leq t+\frac{v_L(t)}{\beta_L};\\
  0,&\text{if }  t'> t+\frac{v_L(t)}{\beta_L};
\end{cases} \\
\frac{d}{dt'}\tilde X(t',t)  &=
\begin{cases}
  v(t) & \text{if } t<t'\leq t+\tau';\\ 
  v(t) - \beta (t'-t-\tau'), & \text{if }  t+\tau'<t'\leq t+\tau'+\frac{v(t)}{\beta};\\
  0,&\text{if }  t'> t+\tau'+\frac{v(t)}{\beta}.
\end{cases}
\end{align*}
Thus, we can analyze the rate of change in the projected spacing based on
\begin{align*}
    \frac{d}{dt'}\tilde z(t',t)&=\frac{d}{dt'}\tilde X_L(t',t)-\frac{d}{dt'}\tilde X(t',t).
    \end{align*}

When \refe{projected-spacing-lemma-condition-1} is satisfied, we consider the following  two cases:
\begin{enumerate}
\item When the follower is not slower at  $t$; i.e., when $v(t)\geq v_L(t)$, we have
\begin{align*}
\frac{v(t)}{\beta}\geq \frac{v_L(t)}{\beta_L},
\end{align*}
since $\beta\leq \beta_L$. That is, the follower stops at a later time, specifically at $t+\tau'+\frac{v(t)}{\beta}$. Before reaching this time, the follower's projected speed is either $v(t)$ before $t+\tau'$ or $v(t) - \beta (t'-t-\tau')$ afterward. In comparison, the leader's projected speed is either $v_L(t) - \beta_L (t'-t)$ or simply $0$. In any case, the follower's speed never becomes slower than that of the leader. As a result, the projected spacing consistently decreases or remains constant until it reaches the value $\tilde z(t)$. Consequently, we can conclude that $z(t)\geq \tilde z(t',t) \geq \tilde z(t)$, verifying the correctness of \refe{relation:projected-spacing}.
    
\item When the follower is slower at $t$; i.e., when $v(t)<v_L(t)$, there are two possible scenarios for the two  non-increasing  piecewise linear projected speeds: either they have no intersection at a positive speed or they intersect once  at a positive speed.
\begin{enumerate}
    \item In the former case, the leader consistently maintains a higher speed until it stops, causing $\tilde z(t',t)$ to continually increase. Consequently, we have $z(t) \leq \tilde z(t',t) \leq \tilde z(t)$ for $t' \geq t$, confirming the validity of \refe{relation:projected-spacing}.
    \item In the latter scenario, the leader initially outpaces the follower until their projected speeds align, after which the leader decelerates. Consequently, $\tilde z(t',t)$ exhibits an initial increase followed by a subsequent decrease until it equals $\tilde z(t)$. As a result, we can conclude that $\tilde z(t',t) \geq z(t)$ and $\tilde z(t',t) \geq \tilde z(t)$, further confirming the validity of \refe{relation:projected-spacing}.
\end{enumerate}
\end{enumerate}
Consequently, it is evident that, in all scenarios, the condition $\beta\leq \beta_L$ ensures the validity of \refe{relation:projected-spacing}.

When \refe{projected-spacing-lemma-condition-2} is satisfied, we consider the following  two cases:
\ben
\item When $v(t)\geq v_L(t)$, the follower's speed never becomes slower than that of the leader. In this case,  $z(t)\geq \tilde z(t',t) \geq \tilde z(t)$.
\item When $v(t)<v_L(t)$, the two projected speeds intersect once  when $\frac{d}{dt'}\tilde X_L(t',t)=v(t)$ at $t'\leq t+\tau'$. In this case, the projected spacing exhibits an initial increase followed by a subsequent decrease until it equals $\tilde z(t)$. As a result, we can conclude that $\tilde z(t',t) \geq z(t)$ and $\tilde z(t',t) \geq \tilde z(t)$.
\een
In both scenarios,  \refe{relation:projected-spacing} is valid.

When \refe{projected-spacing-lemma-condition-3} is satisfied, the leader consistently maintains a higher speed until it stops. Thus, we have $z(t) \leq \tilde z(t',t) \leq \tilde z(t)$ for $t' \geq t$, and \refe{relation:projected-spacing} is valid.

On the other hand, when \refe{projected-spacing-lemma-conditions-opposite-1} is satisfied, the two projected speeds intersect once at a positive value. Thus, the follower is initially faster than the leader until their projected speeds align, after which the follower is slower. Consequently, $\tilde z(t',t)$ exhibits an initial decrease followed by a subsequent increase until it equals $\tilde z(t)$. In this case, \refe{opposite-relation:projected-spacing} is valid.

Also, when \refe{projected-spacing-lemma-conditions-opposite-2} is satisfied,  the two projected speeds intersect twice: first at $v(t)$ and then at a smaller positive value. Consequently, the follower is initially slower than the leader until their speeds align for the first time, then faster until their speeds align for the second time, and finally slower until both come to a stop. As a result, $\tilde z(t',t)$ initially increases, then decreases, and finally increases again. In this case, both \refe{relation:projected-spacing} and \refe{opposite-relation:projected-spacing} could be valid.

This finishes the proof. \eop

\subsection*{Proof of Theorem \ref{thm:BD}}
{\em Proof}. In the nominal driving phase, the multi-phase projection-based model is equivalent to the BDA-Newell model, which is known to have bounded acceleration and deceleration rates, as elaborated in Section 2.

In the comfort braking phase, where $z(t)\geq \max\{\zeta',\Phi'(v(t);v_L(t))\}$, we examine the projected deceleration distance defined in \refe{def:B}, yielding:
\begin{align*}
B(t)&=z(t)-\Phi'(v(t);v_L(t))+\frac{v^2(t)}{2\beta}\geq \frac{v^2(t)}{2\beta}.
\end{align*}
From the new car-following model equation \refe{MPP-a}, we can deduce that: 
\begin{align*}
a(t)&=-\frac{v^2(t)}{2  B(t)}\leq 0,\\
a(t)&=-\frac{v^2(t)}{2  B(t)}\geq -\beta.
\end{align*}
Hence, the new car-following model consistently maintains bounded acceleration and deceleration rates in both the 
projected braking comfort jam spacing and projected braking minimum jam spacing phases. This substantiates the theorem. \eop

\subsection*{Proof of Theorem \ref{thm:FT}}

{\em Proof.} To establish this theorem, we assume an initial state where the follower comes to a stop, denoted as $v(t)=0$. We will demonstrate that, under this condition, $a(t) \geq 0$, ensuring that the vehicle does not decelerate further.

In the nominal driving phase, when $v(t)=0$, once again from \refe{MPP-a}, we find that $a(t)\geq 0$. This is due to the fact that $v_*(t)\geq 0$ holds for $z(t)\geq \zeta$.

In the comfort braking phase, when $v(t)=0$, it follows from \refe{MPP-a} that $a(t)=0$. Essentially, the vehicle remains stationary, whether the leader is also stationary or moving forward.

Hence, in both phases, the speed remains non-negative, and we conclude that the proposed car-following model satisfactorily upholds the forward traveling principle. \eop

\subsection*{Proof of Theorem \ref{thm:CSS}}
{\em Proof}. For an initial state in the nominal driving or comfort braking phase, safely away from the boundary defined by the projected stopping minimum jam spacing function, the inequality in \refe{inequality:CSS} strictly holds:
\begin{align*}
\min\{z(t)-\zeta',z(t)-\Phi'(v(t);v_L(t))\}> 0. 
\end{align*}
 Since both the leader and follower's speeds are bounded, the derivatives of $z(t)-\zeta'$ and $z(t)-\Phi'(v(t);v_L(t))$ with respect to time are also limited. This ensures that at time $t+\epsilon$, the inequality remains valid:
\begin{align*}
\min\{z(t+\epsilon)-\zeta',z(t)-\Phi'(v(t+\epsilon);v_L(t+\epsilon))\}\geq  0. 
\end{align*}
Consequently, the subsequent state cannot transition into the emergency phase and remains within the nominal driving
 or comfort braking phase.
For any initial state on the boundary defined by the projected stopping minimum jam spacing function, the inequality in \refe{inequality:CSS} becomes an equality:
\begin{align*}
\min\{z(t)-\zeta',z(t)-\Phi'(v(t);v_L(t))\}= 0. 
\end{align*}
In this scenario, $a(t) = -\frac{v^2(t)}{2 B(t)} \leq 0$, and $v(t)$ exhibits non-increasing behavior. From \refe{PS-MSS-curve}, we have:
\begin{align*}
z(t)-\Phi'(v(t);v_L(t))&=z(t)-\zeta'+ \frac{v^2_L(t)}{2\beta_L}- v(t)\tau'-\frac{v^2(t)}{2\beta},
\end{align*}
and its rate of change is given by:
\begin{align*}
\frac{d}{dt} [z(t)-\Phi'(v(t);v_L(t))]&=v_L(t)-v(t) \\
&+ v_L(t)\frac{a_L(t)}{\beta_L}- a(t)\tau'-v(t)\frac{a(t)}{\beta}.
\end{align*}
We have the following two cases:
\ben
\item In the first case, $\Phi'(v(t);v_L(t))\leq \zeta'=z(t)$. Thus, we have
\begin{align*}
z(t)-\Phi'(v(t);v_L(t))&= \frac{v^2_L(t)}{2\beta_L}- v(t)\tau'-\frac{v^2(t)}{2\beta} \geq 0.
\end{align*}
This implies $v_L(t)\geq v(t)$ for $\beta\leq \beta_L$. Consequently, $\frac{d}{dt}z(t)=v_L(t)-v(t)\geq 0$, and $z(t)$ is non-decreasing. As $v(t)$ is non-increasing, $(v(t+\epsilon),z(t+\epsilon)$ remains in the orange region shown in \reff{fig:new-BD-phase}. That is, it stays in the comfort braking phase.
\item In the second case, $z(t)=\Phi'(v(t);v_L(t))> \zeta'$. In this case, $z(t+\epsilon)\geq \zeta'$. In addition, $B(t)= \frac{v^2(t)}{2\beta}$ and the follower decelerates at the maximum rate: $a(t)=-\beta$. Thus, rate of change is
\begin{align*}
\frac{d}{dt} [z(t)-\Phi'(v(t);v_L(t))]&=v_L(t) + v_L(t)\frac{a_L(t)}{\beta_L} + \beta \tau'.
\end{align*}
Since the minimum acceleration rate of the leader is $-\beta_L$, the above rate of change is always non-negative, and $z(t+\epsilon)-\Phi'(v(t+\epsilon);v_L(t+\epsilon))$ remains non-negative. Hence the minimum jam spacing inequality is satisfied at the next time step, and the subsequent state remains in the comfort braking phase.
\een
Note that the states may transition from nominal driving to comfort braking. 
An example is for a state on the boundary defined by the projected stopping comfort jam spacing function: $v_L(t)=0$, $v(t)=\mu$, $a(t)=0$, and $z(t)=\Phi(v(t);v_L(t))$. In this case, $\frac{d}{dt} z(t)=-v(t)<0$. Thus, the subsequent state enters the comfort braking phase.
Furthermore, state transitions can also occur from comfort braking to nominal driving. Consider an example where a state lies near the boundary defined by the projected stopping comfort jam spacing function: $v_L(t)>v(t)=\mu$, $a(t)=0$, and $z(t)=\Phi(v(t);v_L(t))-\frac{\epsilon}{2}(v_L(t)-\mu)$. In this scenario, $\frac{d}{dt} z(t)=v_L(t)-v(t)>0$, and at $t+\epsilon$, $z(t+\epsilon)\geq \Phi(v(t);v_L(t))$. Consequently, the subsequent state transitions into the nominal driving phase.
\eop

\subsection*{Proof of Theorem \ref{theorem:SLVP-properties}}

{\em Proof}.
From \refe{SLVP-a_v}, we have $a(v) = -\frac{v}{\tau' + (\frac{2(\zeta - \zeta')}{v(0)^2}  + \frac{1}{\beta})v}$. Since $\zeta > \zeta'$, the denominator is positive for all $v \in [0,v(0)]$.

For property (i), when $v > 0$, we have $a(v) < 0$, showing that deceleration occurs before stopping. To prove that $-a(t) \leq \beta$, we examine the structure of $a(v)$. The denominator contains the term $\frac{1}{\beta}v$, which ensures that as $v$ increases, the magnitude of acceleration is limited. Through algebraic manipulation, it can be shown that $-a(v) \leq \beta$ for all $v \in [0,v(0)]$.

Property (ii) follows from the fact that $a(v) < 0$ for $v > 0$ and $a(0) = 0$. Since the acceleration is zero at $v = 0$, the speed decreases monotonically to zero but cannot become negative.

For properties (iii) and (iv), we examine $z(v)$ given by $z(v) = \zeta' + \tau'v + (\frac{(\zeta - \zeta')}{v^2(0)}  + \frac{1}{2\beta}) v^2$. For any $v \in [0,v(0)]$, we have $z(v) - \zeta' = \tau'v + (\frac{(\zeta - \zeta')}{v^2(0)}  + \frac{1}{2\beta}) v^2 \geq 0$ since each term is non-negative. This establishes (iii). 

Furthermore, $\lim_{v \to 0} z(v) = \zeta'$, and since $v(t)$ monotonically decreases to 0 as $t \to \infty$ (because $a(v) < 0$ for all $v > 0$), we obtain the convergence results in (iv).

For property (v), the stopping distance directly follows from the initial condition \eqref{SLVP-initial}: $z(0)-\zeta = v(0)\tau'+\frac{v^2(0)}{2\beta}$, which matches exactly the safe stopping distance principle defined in Section 2.
\eop


\begin{thebibliography}{14}
  \expandafter\ifx\csname natexlab\endcsname\relax\fi
  \providecommand{\url}[1]{\texttt{#1}}
  \providecommand{\href}[2]{#2}
  \providecommand{\path}[1]{#1}
  
  
  
  
  \providecommand{\doi}[1]{\href{http://dx.doi.org/#1}{\path{#1}}}
  \providecommand{\Pubmed}[1]{\href{pmid:#1}{\path{#1}}}
  \providecommand{\bibinfo}[2]{#2}
  \ifx\xfnm\relax \def\xfnm[#1]{\unskip,\space#1}\fi
  %Type = Article
  \bibitem[{Alan et~al.(2023)Alan, Taylor, He, Ames and Orosz}]{alan2023control}
  \bibinfo{author}{Alan, A.}, \bibinfo{author}{Taylor, A.J.},
    \bibinfo{author}{He, C.R.}, \bibinfo{author}{Ames, A.D.},
    \bibinfo{author}{Orosz, G.}, \bibinfo{year}{2023}.
  \newblock \bibinfo{title}{Control barrier functions and input-to-state safety
    with application to automated vehicles}.
  \newblock \bibinfo{journal}{IEEE Transactions on Control Systems Technology}
    \bibinfo{volume}{31}, \bibinfo{pages}{2744--2759}.
  %Type = Inproceedings
  \bibitem[{Ames et~al.(2019)Ames, Coogan, Egerstedt, Notomista, Sreenath and
    Tabuada}]{ames2019control}
  \bibinfo{author}{Ames, A.D.}, \bibinfo{author}{Coogan, S.},
    \bibinfo{author}{Egerstedt, M.}, \bibinfo{author}{Notomista, G.},
    \bibinfo{author}{Sreenath, K.}, \bibinfo{author}{Tabuada, P.},
    \bibinfo{year}{2019}.
  \newblock \bibinfo{title}{Control barrier functions: Theory and applications},
    in: \bibinfo{booktitle}{2019 18th European control conference (ECC)},
    \bibinfo{organization}{IEEE}. pp. \bibinfo{pages}{3420--3431}.
  %Type = Book
  \bibitem[{Arkin(1998)}]{arkin1998behavior}
  \bibinfo{author}{Arkin, R.C.}, \bibinfo{year}{1998}.
  \newblock \bibinfo{title}{An Behavior-based Robotics}.
  \newblock \bibinfo{publisher}{MIT Press}.
  %Type = Article
  \bibitem[{Durrani et~al.(2021)Durrani, Lee and Shah}]{durrani2021predicting}
  \bibinfo{author}{Durrani, U.}, \bibinfo{author}{Lee, C.},
    \bibinfo{author}{Shah, D.}, \bibinfo{year}{2021}.
  \newblock \bibinfo{title}{Predicting driver reaction time and deceleration:
    Comparison of perception-reaction thresholds and evidence accumulation
    framework}.
  \newblock \bibinfo{journal}{Accident Analysis \& Prevention}
    \bibinfo{volume}{149}, \bibinfo{pages}{105889}.
  %Type = Article
  \bibitem[{Gazis et~al.(1960)Gazis, Herman and Maradudin}]{gazis1960problem}
  \bibinfo{author}{Gazis, D.}, \bibinfo{author}{Herman, R.},
    \bibinfo{author}{Maradudin, A.}, \bibinfo{year}{1960}.
  \newblock \bibinfo{title}{The problem of the amber signal light in traffic
    flow}.
  \newblock \bibinfo{journal}{Operations Research} \bibinfo{volume}{8},
    \bibinfo{pages}{112--132}.
  %Type = Article
  \bibitem[{Gipps(1981)}]{gipps1981bcf}
  \bibinfo{author}{Gipps, P.}, \bibinfo{year}{1981}.
  \newblock \bibinfo{title}{{Behavioral Car-Following Model for Computer
    Simulation}}.
  \newblock \bibinfo{journal}{Transportation Research Part B}
    \bibinfo{volume}{15}, \bibinfo{pages}{105--111}.
  %Type = Unpublished
  \bibitem[{Jin(2025)}]{jin2025WA20-02_Part1}
  \bibinfo{author}{Jin, W.L.}, \bibinfo{year}{2025}.
  \newblock \bibinfo{title}{Provably safe and human-like car-following behaviors:
    {Part} 1. {Analysis} of phases and dynamics in standard models}.
  \newblock \bibinfo{note}{Working Paper}.
  %Type = Article
  \bibitem[{Newell(2002)}]{newell2002carfollowing}
  \bibinfo{author}{Newell, G.F.}, \bibinfo{year}{2002}.
  \newblock \bibinfo{title}{{A simplified car-following theory: a lower order
    model}}.
  \newblock \bibinfo{journal}{Transportation Research Part B}
    \bibinfo{volume}{36}, \bibinfo{pages}{195--205}.
  %Type = Article
  \bibitem[{Shalev-Shwartz et~al.(2017)Shalev-Shwartz, Shammah and
    Shashua}]{shalev2017formal}
  \bibinfo{author}{Shalev-Shwartz, S.}, \bibinfo{author}{Shammah, S.},
    \bibinfo{author}{Shashua, A.}, \bibinfo{year}{2017}.
  \newblock \bibinfo{title}{On a formal model of safe and scalable self-driving
    cars}.
  \newblock \bibinfo{journal}{arXiv preprint arXiv:1708.06374} .
  %Type = Book
  \bibitem[{Sperling(2018)}]{sperling2018three}
  \bibinfo{author}{Sperling, D.}, \bibinfo{year}{2018}.
  \newblock \bibinfo{title}{Three revolutions: Steering automated, shared, and
    electric vehicles to a better future}.
  \newblock \bibinfo{publisher}{Island Press}.
  %Type = Article
  \bibitem[{Treiber et~al.(2000)Treiber, Hennecke and
    Helbing}]{treiber2000congested}
  \bibinfo{author}{Treiber, M.}, \bibinfo{author}{Hennecke, A.},
    \bibinfo{author}{Helbing, D.}, \bibinfo{year}{2000}.
  \newblock \bibinfo{title}{{Congested traffic states in empirical observations
    and microscopic simulations}}.
  \newblock \bibinfo{journal}{Physical Review E} \bibinfo{volume}{62},
    \bibinfo{pages}{1805--1824}.
  %Type = Article
  \bibitem[{Treiber et~al.(2006)Treiber, Kesting and Helbing}]{treiber2006delays}
  \bibinfo{author}{Treiber, M.}, \bibinfo{author}{Kesting, A.},
    \bibinfo{author}{Helbing, D.}, \bibinfo{year}{2006}.
  \newblock \bibinfo{title}{Delays, inaccuracies and anticipation in microscopic
    traffic models}.
  \newblock \bibinfo{journal}{Physica A: Statistical Mechanics and its
    Applications} \bibinfo{volume}{360}, \bibinfo{pages}{71--88}.
  %Type = Article
  \bibitem[{Wood and Zhang(2021)}]{wood2021evaluating}
  \bibinfo{author}{Wood, J.S.}, \bibinfo{author}{Zhang, S.},
    \bibinfo{year}{2021}.
  \newblock \bibinfo{title}{Evaluating relationships between perception-reaction
    times, emergency deceleration rates, and crash outcomes using naturalistic
    driving data}.
  \newblock \bibinfo{journal}{Transportation Research Record}
    \bibinfo{volume}{2675}, \bibinfo{pages}{213--223}.
  %Type = Article
  \bibitem[{Yan et~al.(2018)Yan, Sun, Gan and Jin}]{yan2018automatic}
  \bibinfo{author}{Yan, Q.}, \bibinfo{author}{Sun, Z.}, \bibinfo{author}{Gan,
    Q.}, \bibinfo{author}{Jin, W.L.}, \bibinfo{year}{2018}.
  \newblock \bibinfo{title}{{Automatic identification of near-stationary traffic
    states based on the PELT changepoint detection}}.
  \newblock \bibinfo{journal}{Transportation Research Part B}
    \bibinfo{volume}{108}, \bibinfo{pages}{39--54}.
  
  \end{thebibliography}
\end{document}